\documentclass[journal, twocolumn]{IEEEtran}

\usepackage{cite}
\usepackage{amsmath,amssymb,amsfonts}
\usepackage{algorithmic}
\usepackage{graphicx}
\usepackage{textcomp}
\usepackage{subfigure}
\usepackage{booktabs}
\usepackage{multirow}
\usepackage{makecell}
\def\BibTeX{{\rm B\kern-.05em{\sc i\kern-.025em b}\kern-.08em
		T\kern-.1667em\lower.7ex\hbox{E}\kern-.125emX}}
\usepackage{cuted}
\usepackage{color}
\usepackage{diagbox}
\begin{document}

\title{Centralized Active Reconfigurable Intelligent Surface: Architecture, Path Loss Analysis and Experimental Verification\\
}

\author{Changhao Liu, \IEEEmembership{Student Member, IEEE}, Fan Yang, \IEEEmembership{Fellow, IEEE}, Shenheng Xu, \IEEEmembership{Member, IEEE}, Yezhen Li, and Maokun Li, \IEEEmembership{Senior Member, IEEE}
	\thanks{This work was supported in part by the New Cornerstone Science Foundation through the XPLORER PRIZE. \textit{(Corresponding author: Fan Yang.)}}
	\thanks{The authors are with the Beijing National Research Center for Information Science and Technology (BNRist), Department of Electronic Engineering, Tsinghua University, Beijing 100084, China (e-mail: fan\_yang@tsinghua.edu.cn).}
}

\maketitle

\begin{abstract}
Reconfigurable intelligent surfaces (RISs) are promising candidate for the 6G communication. Recently, active RIS has been proposed to compensate the multiplicative fading effect  inherent in passive RISs. However, conventional distributed active RISs, with at least one amplifier per element, are costly, complex, and power-intensive. To address these challenges, this paper proposes a novel architecture of active RIS: the centralized active RIS (CA-RIS), which amplifies the energy using a centralized amplifying reflector to reduce the number of amplifiers. Under this architecture, only as low as one amplifier is needed for power amplification of the entire array, which can eliminate the mutual-coupling effect among amplifiers, and significantly reduce the cost, noise level, and power consumption. We evaluate the performance of CA-RIS, specifically its path loss, and compare it with conventional passive RISs, revealing a moderate amplification gain. Furthermore, the proposed CA-RIS and the path loss model are experimentally verified, achieving a 9.6 dB net gain over passive RIS at 4 GHz. The CA-RIS offers a substantial simplification of active RIS architecture while preserving performance, striking an optimal balance between system complexity and the performance, which is competitive in various scenarios.
\end{abstract}

\begin{IEEEkeywords}
Active reconfigurable intelligent surface, path loss, reconfigurable intelligent surface (RIS), reconfigurable reflectarray antennas, metasurface.
\end{IEEEkeywords}

\section{Introduction}

\begin{figure}[!t]
	\centerline{\includegraphics[width=1.0\columnwidth]{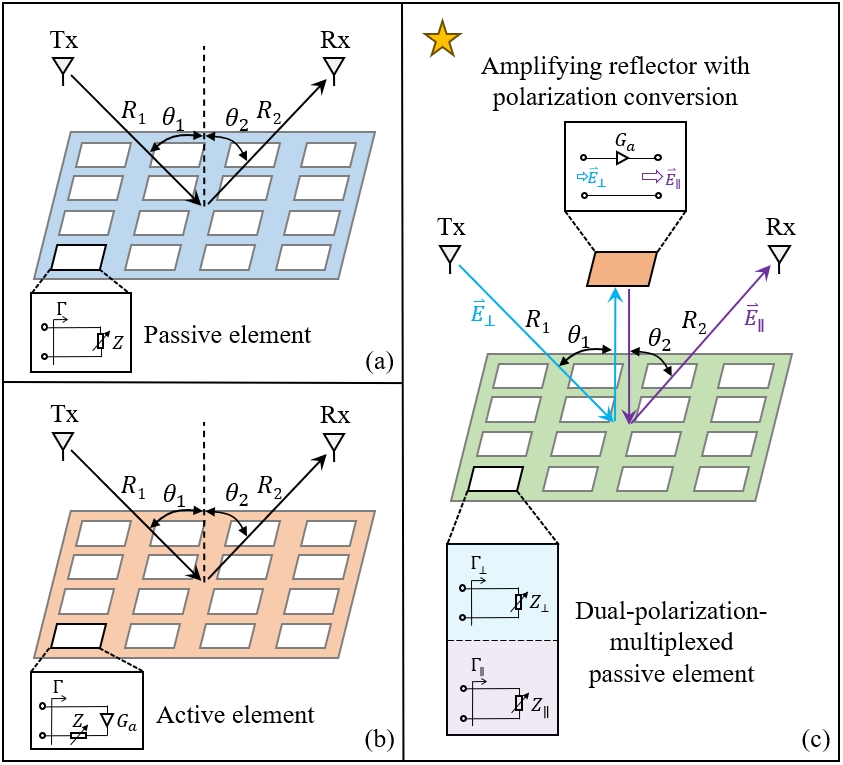}}
	\caption{Architectures of RISs. (a) The passive RIS. (b) The conventional distributed active RIS. (c) The proposed CA-RIS.}
	\label{sche}
\end{figure}

Reconfigurable intelligent surfaces (RISs), comprising numerous phase-reconfigurable metasurface elements, can intelligently control electromagnetic waves in wireless environments, making them a promising candidate for next-generation wireless communications \cite{ris1,ris2}. Recent studies using fabricated passive RISs have experimentally demonstrated dynamic signal enhancement through beam direction scanning \cite{risexp1,risexp2,risexp3,risexp4}. An architecture of passive RIS is shown in Fig. \ref{sche}(a).

While passive RIS can dynamically address blind spots, its signal enhancement in line-of-sight (LOS) scenarios is inherently limited due to the ``multiplicative fading'' effect \cite{aris}, which can double path loss and produce adverse effects. To compensate this multiplicative loss, the concept of active RIS, which adds amplifiers to each element for incoming energy amplification with wave direction tuning, has been proposed \cite{aris}, as depicted in Fig. \ref{sche}(b). Several active RIS designs, demonstrating their power amplifying capabilities, have been developed in \cite{arisexp1,arisexp2,arisexp4,arisexp5,arisexp6,arisexp7}.

However, conventional distributed active RIS designs, featuring at least one amplifier per element, encounter multiple challenges: 
\begin{enumerate}
	\item Increased costs due to additional amplifiers.
	\item Self-excitation and performance degradation from mutual-coupling effects.
	\item Complexity in bias circuitry due to numerous amplifiers.
	\item High power consumption limiting active RIS scalability.
	\item High additive noise of the massive amplifiers.
\end{enumerate}
These physical issues severely limit the total gain and the communication speed of the active RIS. To address these problems, researchers proposed the sub-connected architecture of active RIS \cite{aris2}. By sharing an amplifier across sub-array elements, this architecture can reduce the number of amplifiers by several times. Nevertheless, this sub-connected architecture reduces the energy efficiency, and still requires numerous amplifiers when the number of the active RIS elements becomes large.

To further reduce the number of amplifiers to as low as one, this paper introduces a novel architecture of active RIS: centralized active RIS (CA-RIS), as shown in Fig. \ref{sche}(c). By separating the amplifier from the passive RIS, the incident wave is scattered by the passive RIS towards the centralized amplifying reflector. The reflector, incorporating a single amplifier, amplifies the incident energy with polarization conversion and reflects it to the passive array. The passive RIS then re-directs the wave under the orthogonal polarization to the receiver. This design necessitates only one amplifier for any scale of active RIS, effectively eliminating mutual coupling and drastically reducing costs, complexity, noise figure, and power consumption. Compared with the performance of conventional RISs, path loss analysis and experimental results demonstrate that the CA-RIS achieves a moderate power gain, effectively balancing performance and cost.

This paper is organized as follows: Section II proposes the architecture of CA-RIS, and theoretically analyzes the path loss of CA-RIS, compared to other RIS architectures. Section III details the prototype and performance measurement of CA-RIS, showcasing a 9.6 dB net gain, in line with theoretical analysis. Section IV concludes this study.

\section{The RIS Architectures with Path Loss Analysis}
\label{Sec2}

\subsection{Path Loss Analysis of the Passive RIS} 

\label{secIIA}

The path model of passive RIS is shown in Fig. \ref{sche}(a), where each element can tune the phase of the incident wave from the transmitter (Tx), enabling the RIS to scatter the wave towards the receiver (Rx) with high gain.

A comprehensive discussion of the passive RIS path loss model can be found in \cite{rispath}. For simplicity, it is supposed that the Tx and Rx are both in the far field of RIS, so a modified bistatic radar equation can be applied here \cite{friis}. Let $P_t$ and $G_t$ represent the transmitting power and gain from Tx, respectively, with the distance between Tx and RIS being $R_1$ at an angle of $\theta_1$. Additionally, $G_r$ denotes the gain of Rx's receiving antenna, situated at a distance $R_2$ from the RIS at an angle $\theta_2$. The power received by Rx is given by:
\begin{equation}
	P_r = \frac{P_t G_t}{4\pi R_1^2}\cdot \frac{G_r}{4\pi R_2^2}\cdot A_0^2 \cos(\theta_1) \cos(\theta_2)\cdot \eta_\text{ele}.
	\label{eq7}
\end{equation}
\textit{Proof}: see Appendix A.

Here, $A_0$ represents the physical area of the RIS. 
$\eta_\text{ele}$ is the efficiency of the RIS element, which includes two parts: phase quantization efficiency ($\eta_\text{quant}$) and reflection efficiency ($\eta_\text{loss}$), expressed as $\eta_\text{ele} = \eta_\text{quant}\eta_\text{loss}$. If the element has continuous phase with no reflection magnitude loss, $\eta_\text{ele}=100\%$. Conversely, an element with 1-bit phase quantization and low reflection loss typically incurs a total element loss of about 4-5 dB \cite{phasequant}.

(\ref{eq7}) indicates that the path experiences a multiplicative fading effect, as the product of the two distance terms ($R_1, R_2$) could result in greater path loss than a direct path without RIS, particularly for the passive RISs with small physical areas \cite{aris}.

\subsection{Path Loss Derivation of the Distributed Active RIS}

To compensate the multiplicative fading effect, the active RIS technology is introduced, which incorporates amplifiers into each element alongside the phase shifter, as illustrated in Fig. \ref{sche}(b).

The path loss for an active RIS can be directly deduced from that of a passive RIS. Suppose each amplifier provides a power gain $G_a$, so the formula for the received power at Rx integrates this amplification factor into (\ref{eq7}), resulting in the following expression:
\begin{equation}
	P_r =  G_a \cdot \frac{P_t G_t}{4\pi R_1^2}\cdot \frac{G_r}{4\pi R_2^2}\cdot A_0^2 \cos(\theta_1) \cos(\theta_2)\cdot \eta_\text{ele}.
	\label{eq8}
\end{equation}
This amplifier gain represents the net enhancement of the active RIS over the passive one. Typically, the power received by an RIS in the far field is low, meaning the amplifier processes a weak incident signal. Therefore, the gain of a small-signal amplifier can exceed 20 dB without reaching saturation, effectively compensating for energy losses caused by the multiplicative fading effect.

\subsection{The Proposed CA-RIS with Path Loss Analysis}

\begin{figure}[!t]
	\centerline{\includegraphics[width=0.6\columnwidth]{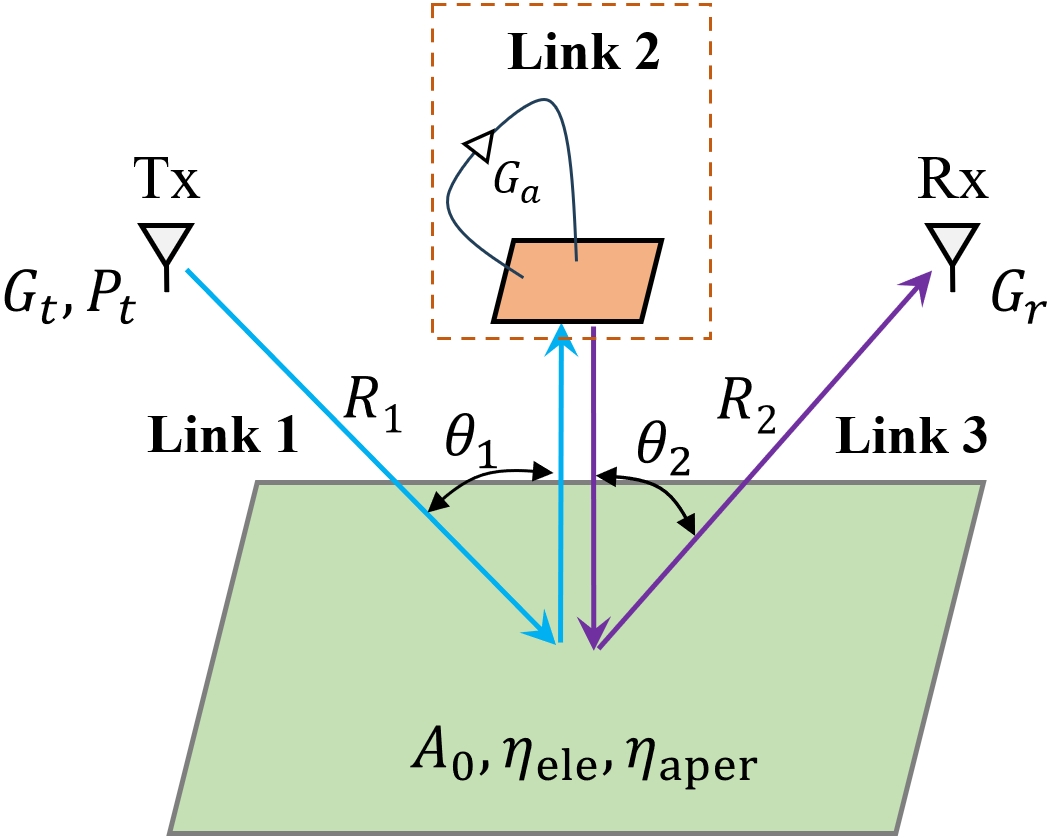}}
	\caption{Illustration of the CA-RIS energy flow links for path loss analysis.}
	\label{CA-RISmodel}
\end{figure}

The conceptual illustration of the proposed CA-RIS is shown in Fig. \ref{sche}(c). The CA-RIS comprises two key components: a dual-polarization-multiplexed passive RIS and an amplifying reflector with polarization conversion, situated in the near field of RIS aperture.  As outlined in Fig. \ref{CA-RISmodel}, the power flow through the CA-RIS involves three distinct stages:
\begin{enumerate}
    \item The incident wave from the transmitter (Tx), with a specific polarization, first interacts with the passive RIS, where it is focused onto the amplifying reflector via phase adjustments by the RIS elements.
    \item The reflector amplifies this incident energy and reflects the augmented energy back to the RIS in an orthogonal polarization.
    \item The RIS receives this reflected wave under orthogonal polarization and directionally re-scatters it towards the receiver (Rx), achieved by coding the phase states of the RIS in this orthogonal polarization.
\end{enumerate}

Consequently, the incident wave is not only amplified and directed towards the Rx, but also has the capability of dynamic beam direction scanning. This architecture distinctly separates the phase-shifting and power-amplifying functions, with energy amplification centralized in a single device. Thus, CA-RIS requires only one amplifier, significantly reducing costs, system complexity, and power consumption. Additionally, in distributed active RIS systems, the noise floor of each amplifier cumulatively increases, leading to substantial noise at the Rx. However, with single-amplifier nature of the CA-RIS, the noise floor of only one amplifier is low. Therefore, the noise figure is considerably reduced in the CA-RIS system.

To evaluate the performance of the proposed CA-RIS, the path loss is analyzed theoretically. The path is divided into three links, as shown in Fig. \ref{CA-RISmodel}. Link 1: from Tx to the centralized reflector; Link 2: within the amplifying circuit of the reflector; Link 3: from the reflector to Rx. 

For better understanding, the path loss of the link 3 is calculated in advance. It is noted that the link 3 is the same as the model of the reflectarray \cite{reflectarray}. This theory holds that reflectarray efficiency comprises two main components: the efficiency of elements ($\eta_\text{ele}$) and the efficiency of the aperture ($\eta_\text{aper}$). The element efficiency, as discussed in Section \ref{secIIA}, combines phase quantization and reflection efficiencies ($\eta_\text{quant}\eta_\text{loss}$). The aperture efficiency encompasses both illumination ($\eta_\text{i}$) and spillover efficiencies ($\eta_\text{s}$), expressed as $\eta_\text{aper} = \eta_\text{i}\eta_\text{s}$. Assuming the aperture's gain at the normal direction is $G_0$, the gain of Link 3 at angle $\theta_2$ is
\begin{equation}
	G_m = G_0 \cos(\theta_2) = \frac{4\pi A_0}{\lambda^2}\eta_\text{ele}\eta_\text{aper}\cdot \cos(\theta_2).
\end{equation}

Here, the reflectarray's feed is the reflector. If the energy transmitted from the reflector is $P_m$, then based on Friis' formula, the power received by Rx is
\begin{equation}
	P_r = \frac{P_m G_m}{4\pi R_2^2}\cdot A_r = P_m \cdot \frac{G_r}{4\pi R_2^2}\cdot A_0\cos(\theta_2)\cdot \eta_\text{ele}\eta_\text{aper}.
	\label{eq10}
\end{equation}
where $A_r$ is the effective aperture area of the receiving antenna, expressed as $A_r = G_r \lambda^2/4\pi$.

Now let us move to the link 1. By reciprocity, we assume that the reflector acts as the transmitter with an emitted power of $\tilde{P}_m$, and the Tx as the receiver with a received power of $P_t$. Therefore, link 1 can also be modeled using the reflectarray theory. Assuming the dual-polarized element has the same efficiency under each polarization, the relationship between $\tilde{P}_m$ and $P_t$ can be similarly written as
\begin{equation}
	P_t = \tilde{P}_m \cdot \frac{G_t}{4\pi R_1^2}\cdot A_0\cos(\theta_1)\cdot \eta_\text{ele}\eta_\text{aper}.
\end{equation}

Given the reciprocal nature of the path, swapping the roles of the transmitter and receiver does not alter the path loss. Hence, if Tx serves as the transmitter with an emitting power of $P_t$, the received power at the reflector is:
\begin{equation}
	\hat{P}_m = P_t \cdot \frac{G_t}{4\pi R_1^2}\cdot A_0\cos(\theta_1)\cdot \eta_\text{ele}\eta_\text{aper}.
	\label{eq12}
\end{equation}

For link 2, suppose the reflector loads the same amplifier as  that in distributed active RIS elements, with an equal amplification gain $G_a$. Consequently, the relationship between the incident power and the reflected power at the reflector is given by:
\begin{equation}
	P_m = G_a \hat{P}_m.
	\label{eq13}
\end{equation}

Finally, by combining equations (\ref{eq10}), (\ref{eq12}), and (\ref{eq13}), the received power at Rx under the CA-RIS architecture is determined as:
\begin{equation}
	P_r = G_a \cdot \frac{P_t G_t}{4\pi R_1^2}\cdot \frac{G_r}{4\pi R_2^2}\cdot A_0^2 \cos(\theta_1) \cos(\theta_2)\cdot \eta^2_\text{ele} \eta^2_\text{aper}.
	\label{eq14}
\end{equation}
This equation reveals that the received power includes the amplifier's gain and is influenced by both the element loss and aperture loss, potentially reducing the overall gain of the CA-RIS.

\subsection{Comparison Among Three RISs}

To effectively compare the performances of various RIS architectures, it is assumed that all parameters, except for the RIS architecture, are identical. Comparing the path loss formulas of the distributed active RIS (\ref{eq8}) and the CA-RIS (\ref{eq14}), the path loss of the CA-RIS ($L^\text{CA}$) exceeds that of the distributed active RIS ($L^\text{DA}$) by a factor of $1/\eta_\text{ele} \eta^2_\text{aper}$, expressed as:
\begin{equation}
	L^\text{CA} = \frac{L^\text{DA}}{\eta_\text{ele} \eta^2_\text{aper} }.
\end{equation}
This factor is solely dependent on element efficiency and aperture efficiency, which can be estimated using reflectarray theory and previous design experiences: For element loss, it is estimated to be around 4 dB for 1-bit phase quantization, approximately 2 dB for 2-bit phase quantization, and about 1 dB for continuous-phase elements. 
For aperture loss, with optimized reflector positioning, the aperture efficiency can exceed 70\% \cite{reflectarray}, corresponding to an aperture loss of roughly 1.5 dB.

\begin{table}[!t]
	\centering
	%\noindent
	\caption{Performance Comparison Among Three RIS Architectures with Numerical Examples}
	\scalebox{0.95}{
		\renewcommand\arraystretch{1.2}
		\setlength\tabcolsep{1mm}{
			\resizebox{\linewidth}{!}{
				\begin{tabular}{llll}
					\hline \hline
					\multicolumn{1}{l}{} & \multicolumn{1}{l}{\makecell{Passive RIS \\ (Baseline)}} & \multicolumn{1}{l}{\makecell{Distributed \\ active RIS}} & \multicolumn{1}{l}{\textbf{CA-RIS}} \\
					\hline
					\makecell[l]{Gain formula} & 1 & \( G_a \) & \(G_a \eta_{\text{ele}} \eta^2_{\text{aper}}  \) \\
					\hline
					\multirow{3}{*}{\makecell[l]{Element loss}} & \multirow{3}{*}{0 dB} & \multirow{3}{*}{0 dB} & 4 dB (1-bit) \\
					&  &  & 2 dB (2-bit) \\
					&  &  & 1 dB (Continuous) \\
					\hline
					\makecell[l]{Aperture loss} & 0 dB & 0 dB & 1.5 dB \\
					\hline
					\makecell[l]{Amplifier gain} & 0 dB & 20 dB & 20 dB \\
					\hline
					\multirow{3}{*}{\makecell[l]{Net gain}} & \multirow{3}{*}{0 dB} & \multirow{3}{*}{20 dB} & 13 dB (1-bit) \\
					&  &  & 15 dB (2-bit) \\
					&  &  & 16 dB (Continuous) \\
					\hline
					\multirow{5}{*}{\makecell[l]{Notes}} & \multirow{5}{*}{-} & \multirow{5}{*}{High gain} & Low complexity \\
&  &  & Low power \\
&  &  & Low cost \\
&  &  & Low noise \\
&  &  & Moderate gain \\
					\hline \hline
			\end{tabular}}
		}
	}
	\label{tab:compare}
\end{table}

Compared to the passive RIS, the received power at Rx using CA-RIS is $ G_a \eta_\text{ele} \eta^2_\text{aper}$ higher. This relationship is expressed as:
\begin{equation}
	L^\text{CA} = \frac{L^\text{pass}}{ G_a\eta_\text{ele} \eta^2_\text{aper}},
\end{equation}
where $L^\text{pass}$ denotes the path loss using the passive RIS. The net gain of the CA-RIS, relative to the passive RIS, is defined as:
\begin{equation}
	G_{\text{net}} = G_a \eta_\text{ele} \eta^2_\text{aper}.
	\label{eqnew10}
\end{equation}

Using the passive RIS as a baseline, Table \ref{tab:compare} provides a straightforward comparison among the three RIS architectures. If the amplifier gain is 20 dB, the net gain can reach up to 16 dB (continuous phase) compared to the passive RIS, and 13 dB under 1-bit phase quantization. Although the CA-RIS has a lower net gain compared to distributed active RIS, its single amplifier design results in lower costs, noise level,  and power consumption. Crucially, without the risk of self-excitation among amplifiers, CA-RIS can utilize high-gain amplifiers (exceeding 20 dB), and be effectively scaled up to massive arrays, greatly enhancing the gain of active RIS. Accordingly, the proposed CA-RIS presents an advantageous trade-off between system complexity and performance, making it a competitive option in various scenarios.

\section{Experimental Verification of CA-RIS}

To validate the CA-RIS concept, a 4-GHz system is built and measured. This section details the construction of the amplifying reflector and the dual-polarization-multiplexed 1-bit RIS, followed by measurements of the path loss of the CA-RIS and its comparison with the passive RIS.

\subsection{Amplifying Reflector Design}
\label{Sec3A}

\begin{figure}[!t]
	\centerline{\includegraphics[width=0.4\columnwidth]{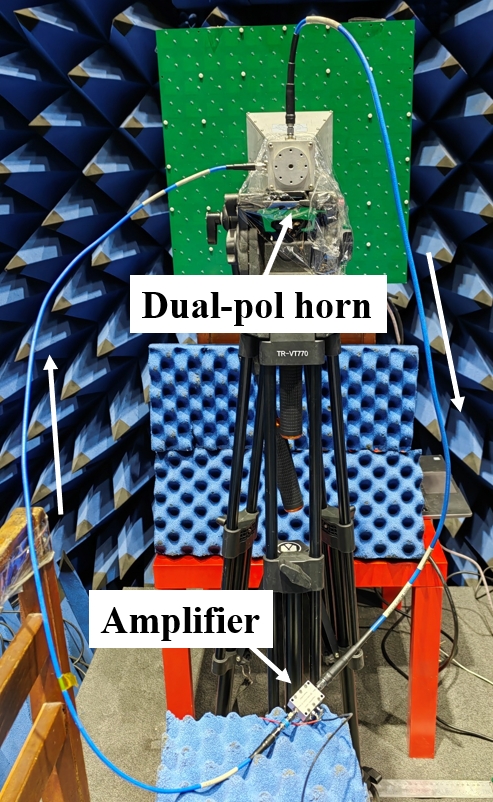}}
	\caption{Design of the amplifying reflector using an amplifier and a dual-polarized horn antenna.}
	\label{reflector}
\end{figure}

To demonstrate the concept, we construct an amplifying reflector. This setup involves a dual-polarized horn antenna linked to a commercial amplifier, as illustrated in Fig. \ref{reflector}. The horn antenna receives the incoming wave with vertical polarization ($Pol_v$), which is then transmitted to the amplifier via a cable. After amplification from the amplifier, the wave is emitted from another port of the horn antenna with horizontal polarization ($Pol_h$), thereby achieving energy amplification with polarization conversion.

\begin{figure}[!t]
	\centerline{\includegraphics[width=0.5\columnwidth]{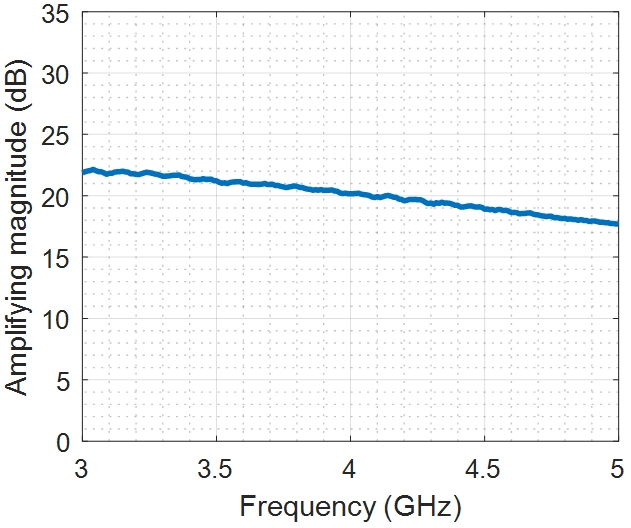}}
	\caption{Measured gain of the amplifier.}
	\label{ampS21}
\end{figure}

A commercial low noise amplifier (LNA) functional across 20 MHz - 6 GHz, is chosen as the centralized amplifier.
The amplification gain of the LNA (connected with an attenuator for protection) is measured using a vector network analyzer (VNA), and the results are shown in Fig. \ref{ampS21}. At 4 GHz, the LNA shows a gain of 20.1 dB, with a stable performance within its operational bandwidth. When connected with the horn, it can amplify spatial waves by approximately 20.1 dB with polarization conversion.

The dual-polarized horn antenna JXTXLB-SJ-20180, operating between 2 - 18 GHz, has a $Q_f$ factor of 2.5 at 4 GHz. Following reflectarray theory \cite{reflectarray}, the position of the feed should be optimized to maximize $\eta_\text{aper}$. For this setup, the optimal F/D ratio is 0.5, resulting in a simulated aperture loss of 1.6 dB.

It is important to note that while our amplifying reflector uses commercial components, fabricating an integrated reflection amplifier with polarization conversion is feasible \cite{amppatch} and potentially more compact than our demonstration model.

\subsection{Dual-Polarization-Multiplexed RIS}
\label{Sec3B}

\begin{figure}[!t]
	\centerline{\includegraphics[width=0.5\columnwidth]{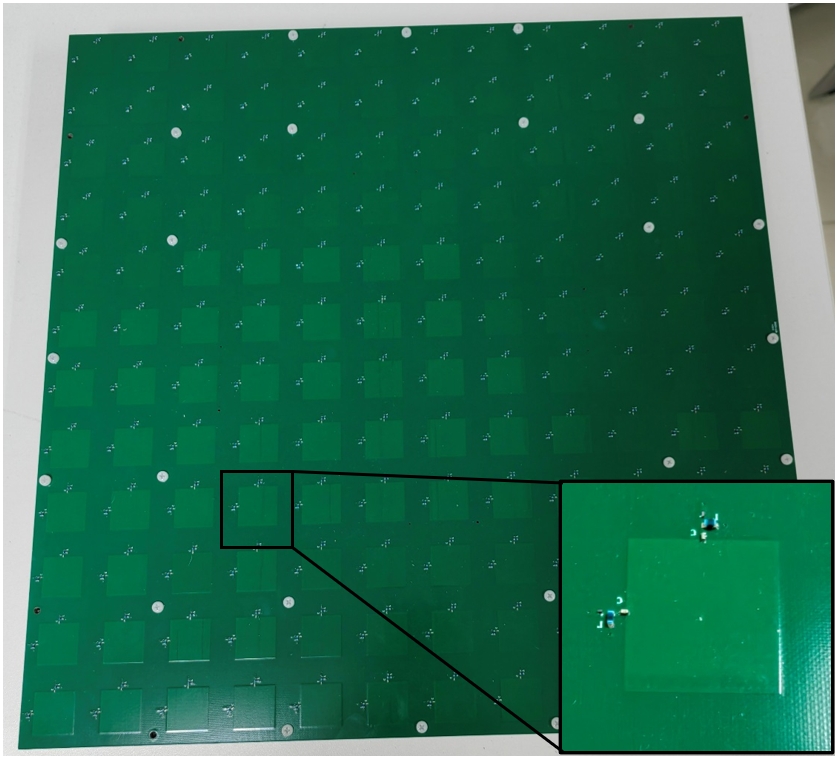}}
	\caption{Illustration of the dual-polarized passive 1-bit RIS prototype.}
	\label{dualpolRIS}
\end{figure}

The development of dual-polarization-multiplexed passive RIS hardware is also feasible, as evidenced by various prototypes \cite{polris1, polris2,polris3,polris4}. For demonstration, we utilize a 4-GHz commercial prototype from Actenna Technology, as depicted in Fig. \ref{dualpolRIS}. This 1-bit RIS array consists of 12 $\times$ 12 elements, with an overall size of 405 mm $\times$ 405 mm. Each element is equipped with two orthogonal PIN diodes for independent phase control at both polarizations.

\begin{figure}[!t]
	\centering
	\subfigure[] { 
		\includegraphics[width=0.45\columnwidth]{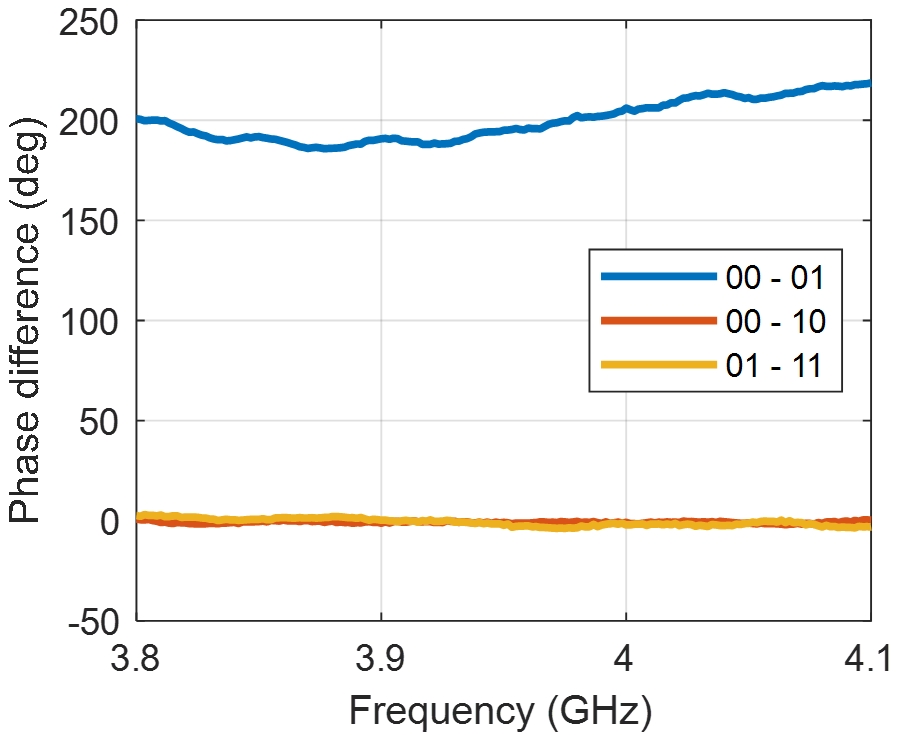} 
	} 
	\subfigure[] { 
		\includegraphics[width=0.45\columnwidth]{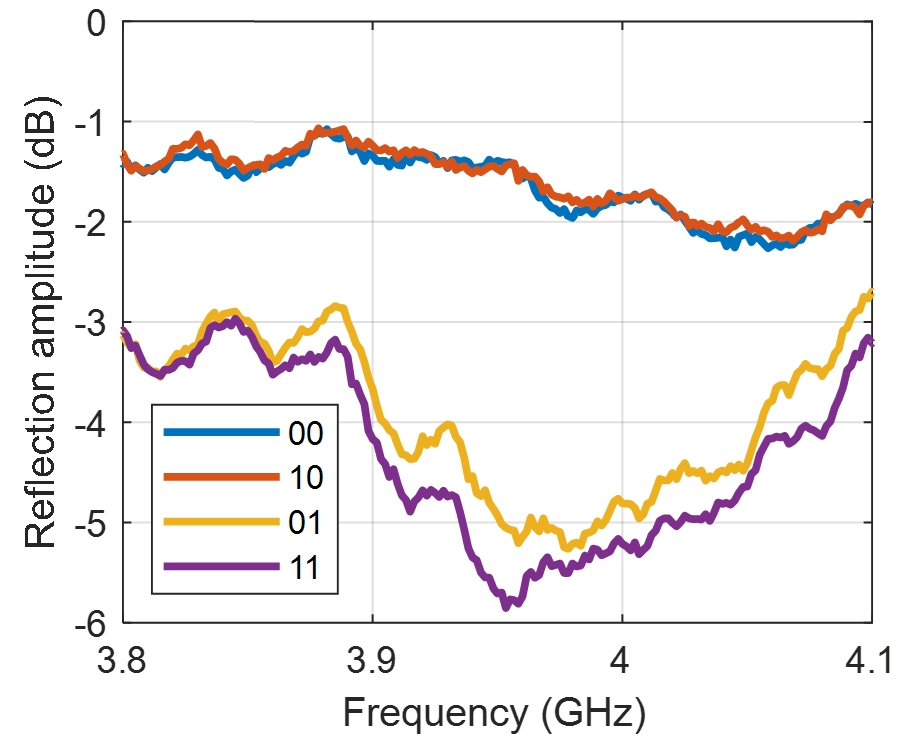} 
	} 
		\subfigure[] { 
		\includegraphics[width=0.45\columnwidth]{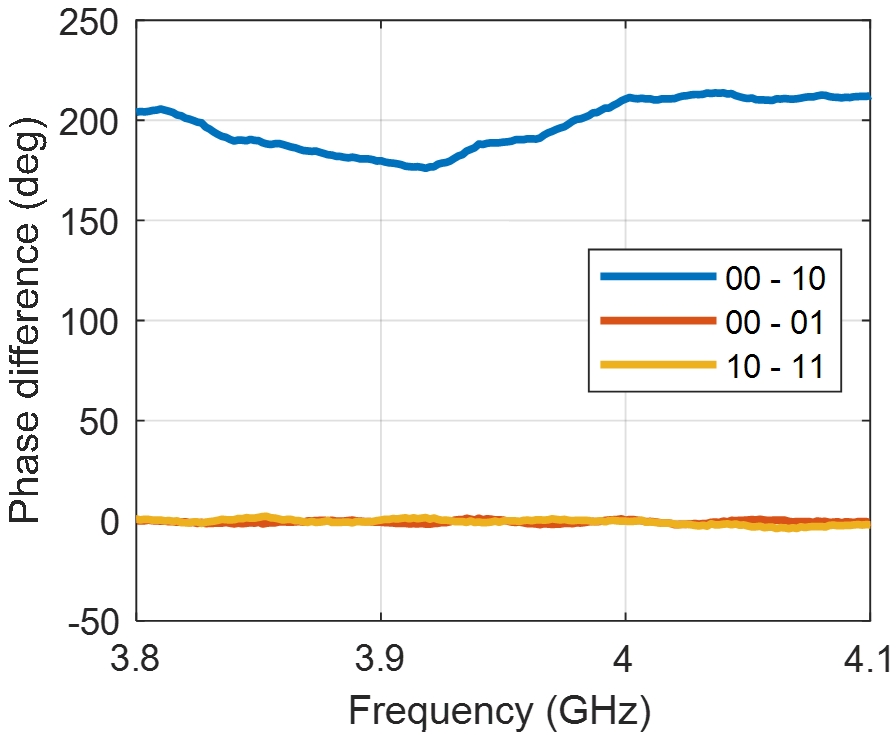} 
	} 
	\subfigure[] { 
		\includegraphics[width=0.45\columnwidth]{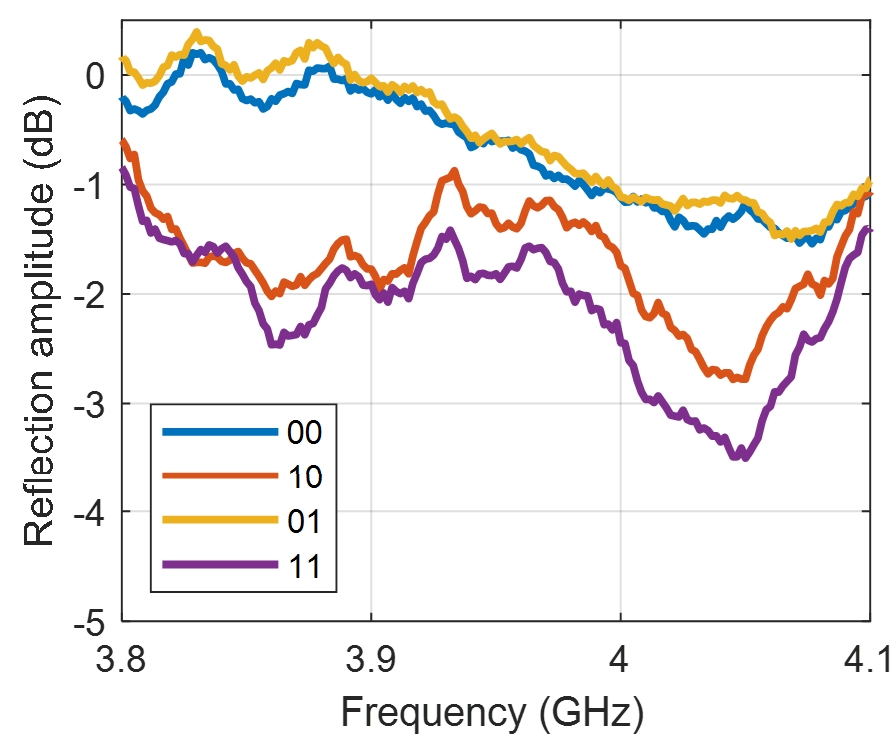} 
	} 
	\caption{Measurement results of the passive RIS. (a) Reflection phase differences under $Pol_h$. (b) Reflection amplitudes under $Pol_h$. (c) Reflection phase differences under $Pol_v$. (d) Reflection amplitudes under $Pol_v$.}
	\label{RISmeas}
\end{figure}

The reflection phase and amplitude performance of the RIS array are measured using the far-field scattering method. The relationship between the PIN diode states and their corresponding codes is summarized in Table \ref{relation}, with PIN$_h$ and PIN$_v$ denoting the horizontal and vertical PIN diodes, respectively. The measured results are presented in Fig. \ref{RISmeas}. For $Pol_h$, the reflection phase differences from 3.8 GHz to 4 GHz are within 180$^\circ\pm$25$^\circ$, and the average reflection amplitude is around -3.3 dB at 4 GHz. Additionally, altering the PIN$_v$ codes does not impact the phase and amplitude of the $Pol_h$ reflected wave. Similarly, for $Pol_v$, the phase differences are also approximately 180$^\circ$, with an average reflection amplitude of -1.7 dB at 4 GHz for all four states. Changes in PIN$_h$ states similarly show minimal effect on $Pol_v$ states. Therefore, the RIS operates effectively from 3.8 GHz to 4 GHz, with an average reflection amplitude loss around 2.5 dB. Along with the 3 dB loss due to 1-bit phase quantization, the total element loss is approximately 2.5 + 3 = 5.5 dB.

\begin{table}[!t]
	\centering
	%\noindent
	\caption{Relation Between PIN Diode States and the Codes}
	%\scalebox{0.95}{
		\renewcommand\arraystretch{1.2}
		\setlength\tabcolsep{1mm}{
			{
				\begin{tabular}{|c|c|c|}
					\hline
					\diagbox{PIN$_v$}{PIN$_h$} & OFF & ON \\ \hline
					OFF & 00 & 01\\ \hline
					ON & 10 & 11\\ \hline
				\end{tabular}
		}}
		\label{relation}
	\end{table}

\subsection{Measurement of CA-RIS}

 \begin{figure*}[!t]
	\centerline{\includegraphics[width=1.8\columnwidth]{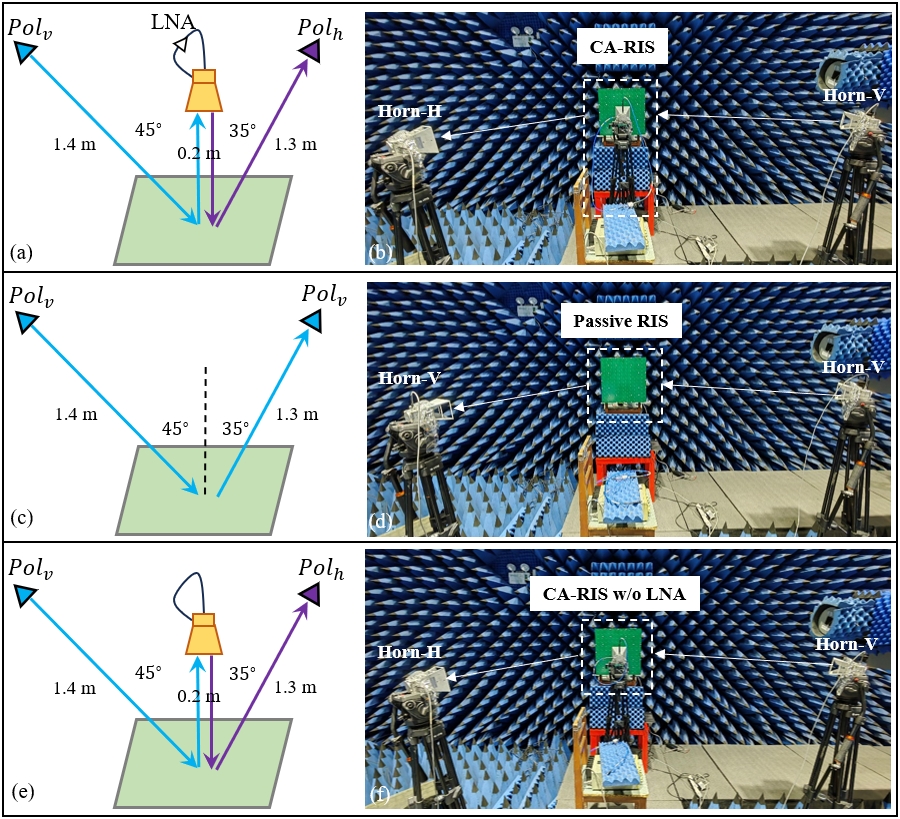}}
	\caption{Measurement schematics and setups of (a)-(b) the CA-RIS, (c)-(d) the passive RIS, and (e)-(f) the CA-RIS without the LNA.}
	\label{meassetup}
\end{figure*}

A series of measurements are conducted to validate the CA-RIS design and the path loss relation between CA-RIS and passive RIS. The experimental setups are shown in Fig. \ref{meassetup}. For optimal $\eta_\text{aper}$, the focal distance of the horn is set to 0.2 m. To minimize the horn blockage effect, the incident and reflected angles are set to 45$^\circ$ and 35$^\circ$, respectively. Given the space constraints of the microwave chamber and the VNA's transmitting power, the distances from Tx and Rx to the RIS are 1.4 m and 1.3 m, respectively, approximating the far field. All three horns are aligned at the same height, corresponding to the central height of the RIS array. 

\begin{figure}[!t]
	\centering
	\subfigure[] { 
		\includegraphics[width=0.4\columnwidth]{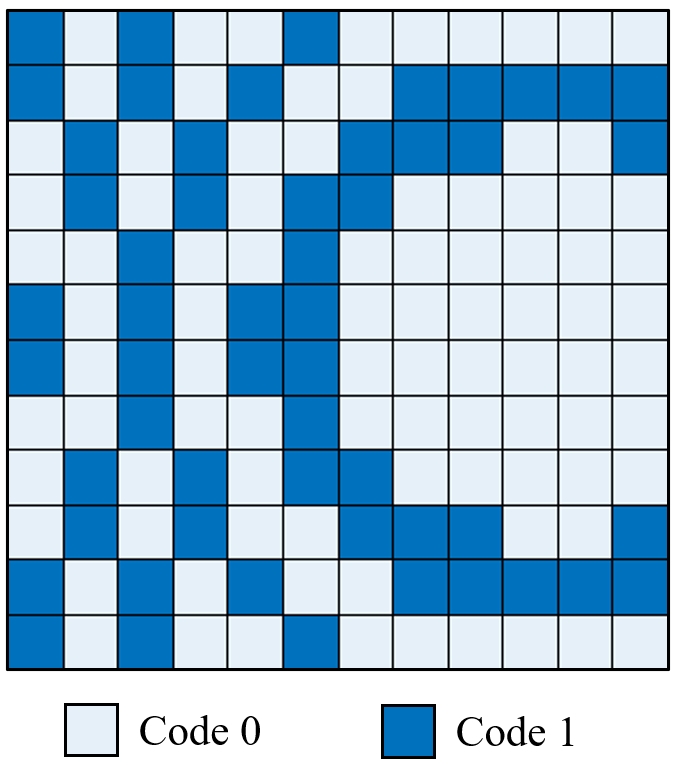} 
	} 
	\subfigure[] { 
		\includegraphics[width=0.4\columnwidth]{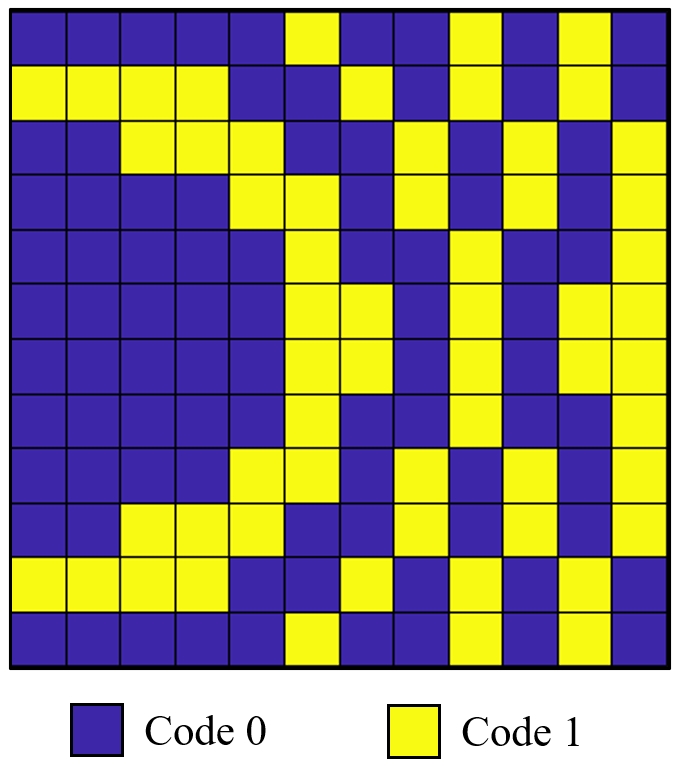} 
	} 
	\caption{Coding patterns of the CA-RIS. (a) Coding pattern for the vertical diodes, corresponding to the 35$^\circ$ incident wave. (b) Coding pattern for the horizontal diodes, corresponding to the 45$^\circ$ reflected wave.}
	\label{coderelay}
\end{figure}

\begin{figure}[!t]
	\centerline{\includegraphics[width=0.6\columnwidth]{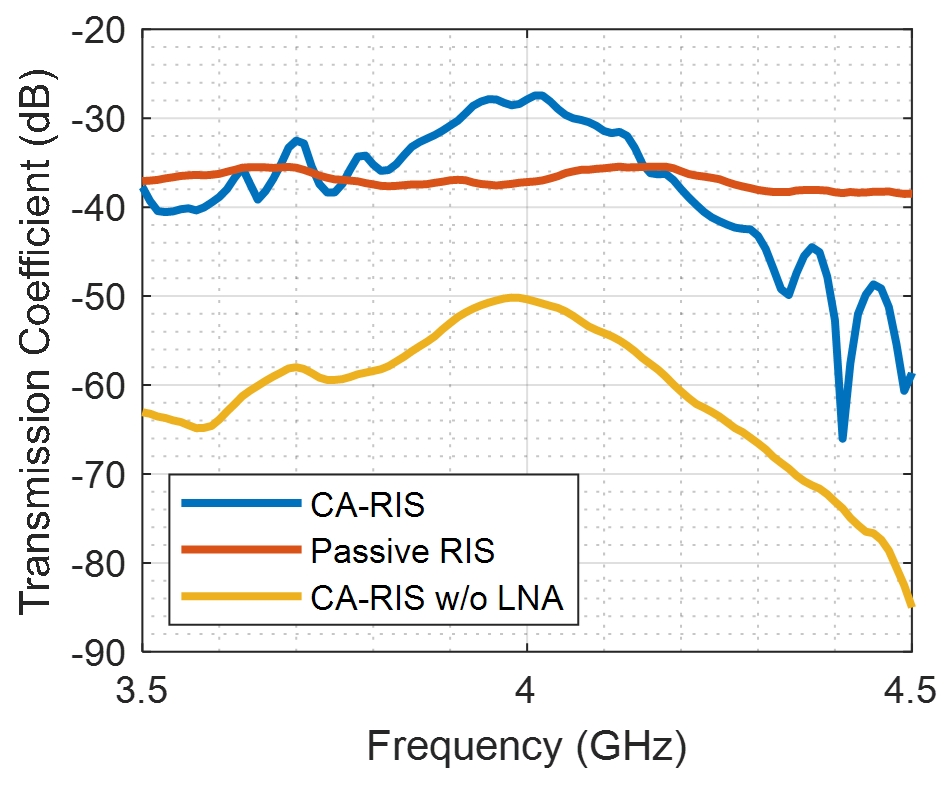}}
	\caption{Measured transmission coefficients of the CA-RIS, passive RIS and CA-RIS without LNA respectively.}
	\label{measres}
\end{figure}

The measurement setup of CA-RIS is shown in Fig. \ref{meassetup}(a)-(b). A vertical-polarized feed horn (Tx) directs at the center of the CA-RIS. After energy amplification and polarization conversion from the centralized amplifier, horizontally polarized amplified wave is reflected towards the Rx, received by a horizontally polarized horn. The coding patterns for the RIS are designed based on the reconfigurable reflectarray method \cite{reflectarray}, treating the centralized horn as the feed and both Rx and Tx as receivers. These patterns, depicted in Fig. \ref{coderelay}, generate beams with orthogonal polarizations pointing at Rx and Tx, respectively. The measured transmission coefficient (Fig. \ref{measres}) for the CA-RIS shows a peak of -27.5 dB at 4.01 GHz. Around this central frequency, the transmission magnitude decreases on either side, with the 3 dB bandwidth extending from 3.91 GHz to 4.08 GHz.

\begin{figure}[!t]
	\centerline{\includegraphics[width=0.4\columnwidth]{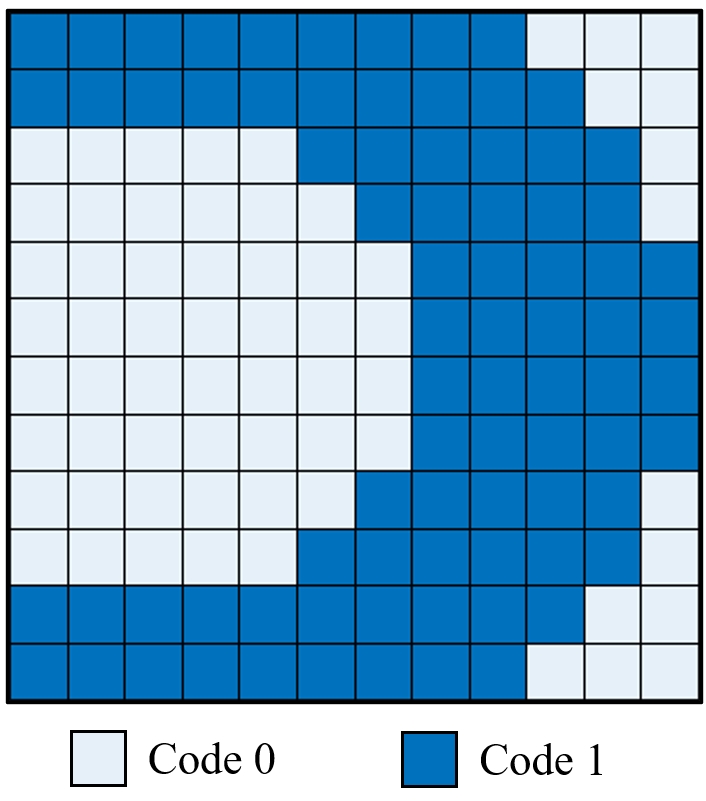}}
	\caption{Coding pattern of the passive RIS under vertical polarization.}
	\label{coderis}
\end{figure}

Comparative measurements are also made with the passive RIS, using the same RIS hardware but only employing one polarization channel ($Pol_v$), as illustrated in Fig. \ref{meassetup}(c)-(d). The coding pattern of passive RIS under the vertical polarization is shown in Fig. \ref{coderis}, directing the incident vertical wave towards Rx. The measured result is shown in Fig. \ref{measres}. At 4.01 GHz, the passive RIS exhibited a transmission amplitude of -37.1 dB, indicating a 9.6 dB net gain of the CA-RIS over the passive RIS. Moreover, the CA-RIS shows net gain benefits in the range of 3.76 GHz to 4.15 GHz.

An additional experiment is conducted to confirm that the net gain results from the LNA amplification rather than the horn or holder. This involves removing the LNA while retaining the same horns and codes, as shown in Fig. \ref{meassetup}(e)-(f). This setup is illustrative only, as the measured received power is below the system's cross-polarization level. Therefore, the result may not represent the real received power. Consequently, the actual results shown in Fig. \ref{measres} are derived from a two-stage measurement process detailed in Appendix B. The CA-RIS without LNA shows a maximum transmission coefficient of -50.2 dB at 3.98 GHz. Comparing this with the CA-RIS curve, the difference with and without LNA is 22.7 dB at 4.01 GHz, aligning with the amplifying gain of the LNA. Note that the LNA gain is 20.1 dB, and the slight discrepancy can be attributed to the repeated cable loss accounted during the measurement without the LNA. Furthermore, the similarity in the trends of both curves is evident, with the CA-RIS curve exhibiting an entire elevation of approximately 22 dB over the curve without the LNA. The consistency in these measurements reinforces the conclusion that the net gain is indeed a result of the LNA's amplification. 

In summary, the experimental results successfully demonstrate the CA-RIS concept, showing a significant net gain of 9.6 dB around 4 GHz compared to the passive RIS. The comparative experiment, revealing a 22 dB increase with the inclusion of the LNA, closely matches the LNA's amplification gain, thus confirming the correctness of the CA-RIS measurement.

\subsection{Discussions}

To confirm the analysis conducted in Section \ref{Sec2}, especially (\ref{eqnew10}), a relative path loss budget for the CA-RIS at 4 GHz is summarized in Table \ref{pathloss}. The measured gain of the amplifier is 20.1 dB. After accounting for the element loss and the aperture loss as discussed in Sections \ref{Sec3A} and \ref{Sec3B}, the ideal gain reduces to 11.4 dB. Additionally, due to the relatively large size of the horn compared to the aperture, which may block the energy flow, losses attributed to horn blockage and other factors are estimated to be 1.8 dB. Consequently, the net gain is calculated to be 9.6 dB, aligning with the measured results.

\begin{table}[!t]
	\centering
	\caption{Relative Path Loss Budget for the CA-RIS at 4 GHz}
	\renewcommand\arraystretch{1.2}
	\begin{tabular}{ll}
		\hline
		\hline
		Theoretical formula & $G_{\text{net}}=G_a\eta_{\text{ele}}\eta^2_{\text{aper}} $ \\
		Amplifier gain & 20.1 dB \\
		Element loss & 5.5 dB \\
		Aperture loss & 1.6 dB \\
		Horn blockage and others & 1.8 dB \\
		Measured net gain & 9.6 dB \\ \hline \hline
	\end{tabular}
	\label{pathloss}
\end{table}

Comparing this with Table \ref{tab:compare}, the theoretical net gain for a 1-bit system should be 13 dB when the amplifier gain is 20 dB. This discrepancy suggests there is potential to further improve the experimental net gain. Potential avenues for improvement may include:
\begin{itemize}
	\item Reducing element loss: This could be achieved by using low-loss switches and enhancing the design of the RIS elements.
	\item Customized amplifying reflector: Designing an integrated amplifying reflector with a smaller footprint could minimize horn blockage of energy flow.
	\item Adopting transmissive RIS: Transmissive RIS not only has better conformal characteristics, but can also eliminate the horn blockage.
	\item Scaling up the RIS array: Increasing the size of the RIS array can not only increase the absolute gain but also reduce the relative effect of horn blockage.
\end{itemize}

A comparison with the measurement results of other distributed active RIS is summarized in Table \ref{compare}. Notably, all the active RISs in the comparison operate under 6 GHz. This study is unique in its use of a centralized architecture, which separates the amplifier from the passive RIS. In typical distributed active RIS systems, the bandwidth is constrained by the amplifier's circuitry. However, in the case of the CA-RIS, the bandwidth limitation is governed by the passive RIS. This limitation can be significantly mitigated by employing a broadband passive RIS, thus potentially enhancing the bandwidth of the CA-RIS beyond that of conventional distributed active RISs. Besides, the centralized architecture effectively reduces the number of required amplifiers to just one, a significant reduction compared to the distributed systems in other research. The size of the RIS in this work, 12$\times$12, is the largest among others, which can increase its absolute gain and enhance its capability for beam manipulation. Additionally, this centralized design allows for an expansion of the array size without needing more amplifiers. The net gain achieved in this study is 9.6 dB, a notable figure that suggests a high capability for signal enhancement. This net gain also has the potential to be further increased by using an amplifier with a higher gain. Thus, the proposed CA-RIS in this work offers a beneficial balance between system complexity and performance. It shows significant potential for enhancing performance without adding to system complexity, making it a potentially more efficient and effective approach in the application of RIS technologies.

\begin{table}[!t]
	\centering
	\caption{Comparison with Other Distributed Active RISs}
	\renewcommand\arraystretch{1.1}
	\setlength\tabcolsep{1mm}{
				\resizebox{\linewidth}{!}{
			\begin{tabular}{lcccc}
				\hline
				\hline
				Literature &  Frequency (GHz) & Amplifier number & RIS size & Net Gain (dB)                  \\ \hline
				\cite{arisexp1}  & 5.7 & 48 & 6$\times$8 & 10.0  \\
				\cite{arisexp4} & 3.5 &64 & 8$\times$8 & 10 \\
				\cite{arisexp5} & 2.4 & 4 & 2$\times$2 & 7-11  \\
				\cite{arisexp6}  & 4.0 & 100 & 10$\times$10 & 7-9  \\
				\cite{arisexp7}  & 3.45-3.54 & 32& 4$\times$4 & 6.7-8  \\
				\textbf{This work} & \textbf{3.91-4.08} & \textbf{1} & \textbf{12$\times$12} & \textbf{9.6} \\
				\hline \hline
			\end{tabular}}
	}
	\label{compare}
\end{table}	

% comparison among different active RISs

\section{Conclusion}

The novel CA-RIS architecture for active RIS significantly reduces the number of amplifiers by utilizing a centralized amplifying reflector. By using as low as only one amplifier per array, this approach not only lowers fabrication costs and reduces energy consumption, but also simplifies the system and mitigates mutual-coupling effects. Furthermore, the low-noise and wide bandwidth nature of CA-RIS can potentially improve the communication speed. Path loss analysis reveals that CA-RIS achieves a moderate net gain exceeding 13 dB over passive RIS. Experimental validation confirms CA-RIS's viability with a 9.6 dB net gain at 4 GHz. Overall, CA-RIS strikes an optimal balance between performance and complexity, marking it as a promising solution for active RIS applications.

Future enhancements for CA-RIS include employing larger RIS arrays to boost the directivity, implementing 2-bit RIS to decrease the phase quantization loss, using higher-gain amplifiers to significantly elevate net gain, and developing transmissive RIS to eliminate the horn blockage.

\section*{Appendix A: Path Loss Analysis of Passive RIS}

As shown in Fig. \ref{sche}(a), suppose the emitting power and gain from the Tx are $P_t$ and $G_t$ respectively, and the distance between the Tx and RIS is $R_1$. Hence, the power received by the RIS is
\begin{equation}
	P_s = \frac{P_t G_t}{4\pi R_1^2}\cdot A_s,
	\label{eq1}
\end{equation}
where $A_s$ is the effective aperture area of the RIS at the oblique angle $\theta_1$. Simply suppose the radiation pattern of the RIS is modeled by $\cos(\theta)$ function, so $A_s$ equals to 
\begin{equation}
	A_s = A_0 \cos(\theta_1),
\end{equation}
where $A_0$ is the physical area of the RIS.

Then the energy is scattered by the RIS toward Rx, and similarly, the received power at the Rx is
\begin{equation}
	P_r = \frac{P_s G_s}{4\pi R_2^2}\cdot A_r.
	\label{eq3}
\end{equation}
In (\ref{eq3}), $A_r$ is the effective aperture area of the receiving antenna, and the expression is
\begin{equation}
	A_r = \frac{G_r \lambda^2}{4\pi},
\end{equation}
where $G_r$ is the gain of the receiving antenna at Rx.
$G_s$ is the emitting gain of the RIS at angle $\theta_2$, which is expressed as
\begin{equation}
	G_s = G_0 \cos(\theta_2).
\end{equation}
$G_0$ is the gain of the RIS at normal angle, which is
\begin{equation}
	G_0 = \frac{4\pi A_0}{\lambda^2}\eta_\text{ele}.
	\label{eq6}
\end{equation}
$\eta_\text{ele}$ is the efficiency of the RIS element, which includes phase quantization efficiency  and reflection efficiency.

Rearranging (\ref{eq1})-(\ref{eq6}), the received power at Rx is obtained
\begin{equation}
	P_r = \frac{P_t G_t}{4\pi R_1^2}\cdot \frac{G_r}{4\pi R_2^2}\cdot A_0^2 \cos(\theta_1) \cos(\theta_2)\cdot \eta_\text{ele}.
	\label{Aeq7}
\end{equation}

\section*{Appendix B: The Two-Step Measurement of the CA-RIS Without LNA}

\begin{figure}[!t]
	\centering
	\subfigure[] { \label{S2stp1}
		\includegraphics[width=0.45\columnwidth]{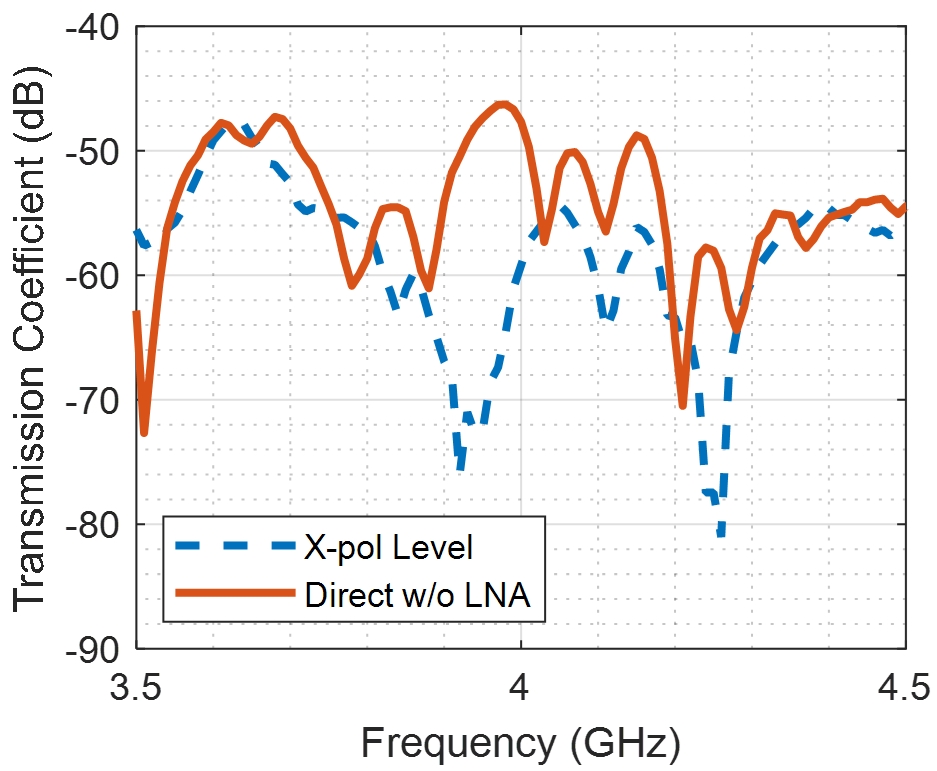} 
	} 
	\subfigure[] { \label{S2stp2}
		\includegraphics[width=0.45\columnwidth]{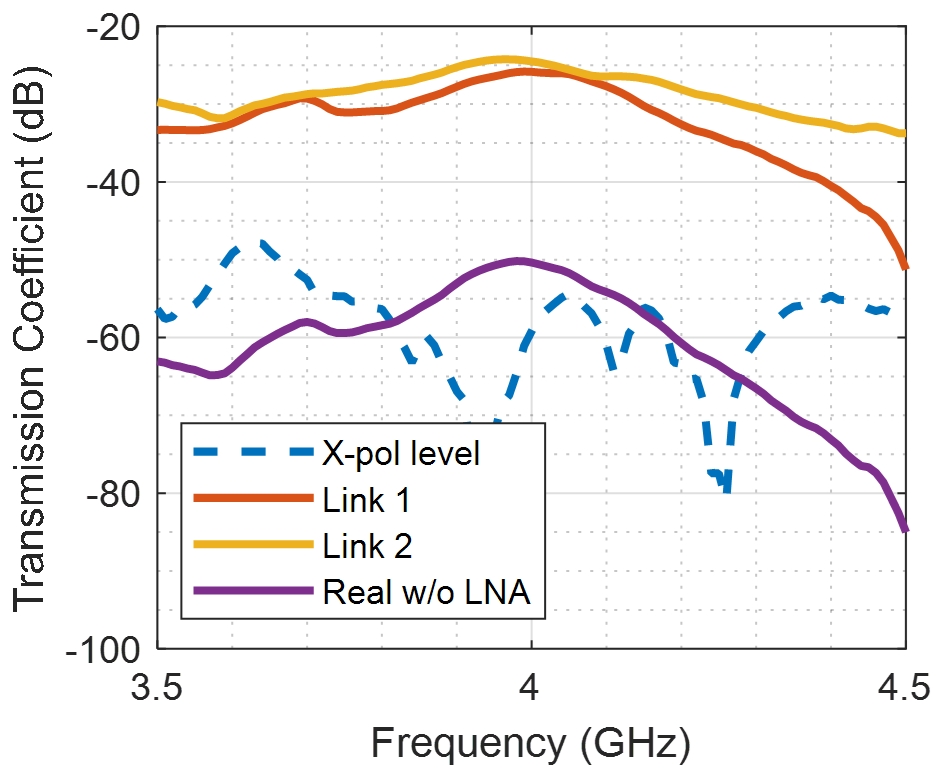} 
	} 
	\caption{The direct measured result and the two-step measured result. (a) Measured cross-polarization level of the system and the direct measured resuls based on the setup in Fig. \ref{meassetup}(e)-(f). (b) Results based on the two-step measurement. The real path loss without LNA is obtained by summing two measured link losses, which is not affected by the cross polarization level.}
	\label{S2stp}
\end{figure}

The CA-RIS without LNA, as depicted in Fig. \ref{meassetup}(e)-(f), presents a measurement challenge: the high path loss in the link results in the cross-polarization level in the system comparable to the measured transmission coefficient, as seen in Fig. \ref{S2stp1}. Consequently, the direct measurement does not accurately reflect the actual transmission coefficient.

\begin{figure}[!t]
	\centerline{\includegraphics[width=0.7\columnwidth]{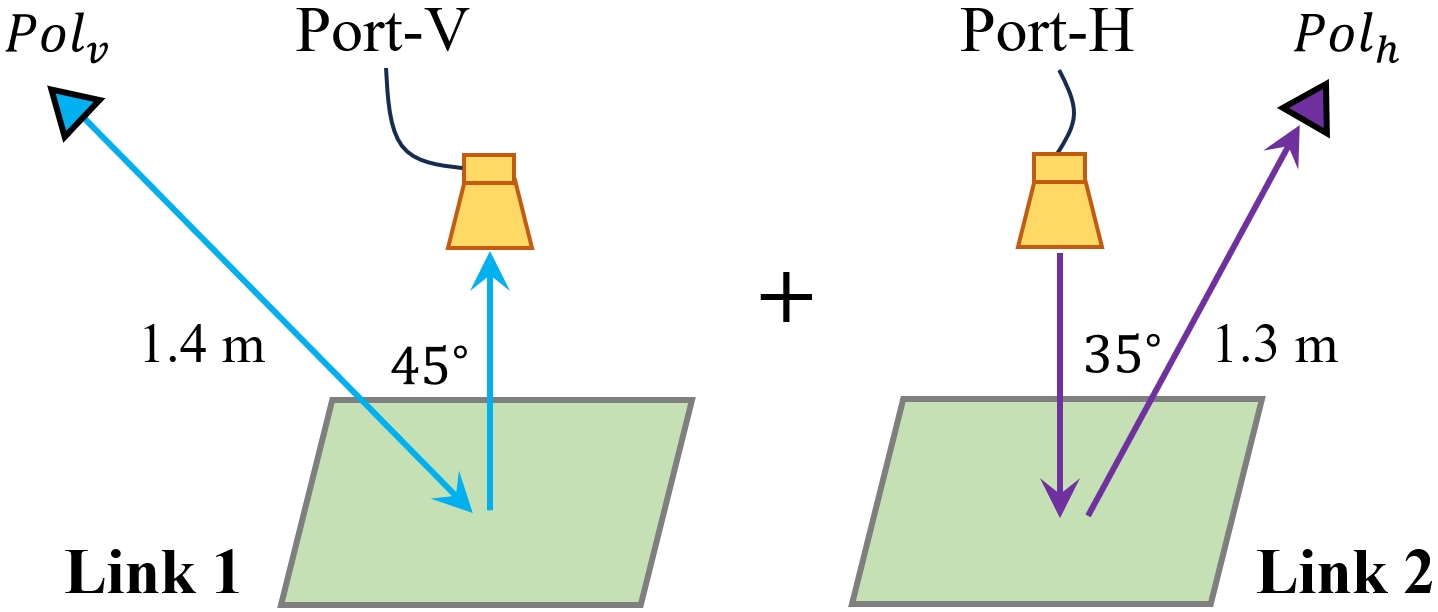}}
	\caption{Schematic of the two-step measurement.}
	\label{f2stp}
\end{figure}

To address this issue, we implement a two-step measurement approach. This method involves dividing the total path into two separate links, each with a transmission coefficient above the system's cross-polarization level, as illustrated in Fig. \ref{f2stp}. By measuring each link individually and then summing the two measured link losses, the real path loss can be accurately determined.

Fig. \ref{S2stp2} demonstrates that the transmission coefficients in each link exceed the system's cross-polarization level. Therefore, by adding these two coefficients, the actual transmission coefficient of the CA-RIS without LNA is successfully obtained.

\end{document}